\documentclass{aa}  
\usepackage{graphicx}
\usepackage{amssymb}
\usepackage{txfonts}
\usepackage{mathrsfs}
\usepackage{array}
\usepackage{natbib}
\bibpunct{(}{)}{;}{a}{}{,}   
\begin{document}
   \title{Reddening and metallicity maps of the Milky Way Bulge from VVV and 2MASS\thanks{Based on observations taken within the ESO VISTA Public Survey VVV, Programme ID 179.B-2002}}
   \subtitle{I. The method and minor axis maps}
   
   \author{O. A. Gonzalez$^{1}$  \and M. Rejkuba$^{1}$ \and M. Zoccali$^{2}$ \and E. Valenti$^{1}$ \and D. Minniti$^{2,3,4}$}
   
   \offprints{O. A. Gonzalez}
   \institute{ $^{1}$European
   Southern Observatory, Karl-Schwarzschild-Strasse 2, D-85748 Garching,
Germany\\ \email{ogonzale@eso.org; mrejkuba@eso.org; evalenti@eso.org}\\
   $^{2}$Departamento    Astronom\'ia    y Astrof\'isica,
   Pontificia Universidad  Cat\'olica de Chile,  Av. Vicu\~na Mackenna
   4860,         Stgo.,         Chile\\         \email{mzoccali@astro.puc.cl;
Dante@astro.puc.cl}\\
   $^{3}$Vatican Observatory, V00120 Vatican City State, Italy\\
   $^{4}$European Southern Observatory, Ave. Alonso de Cordova 3107, Vitacura, Santiago, Chile\\
}
   \date{Received ; Accepted }

  
\abstract  
{The Milky Way bulge is the nearest galactic bulge and the best laboratory for studies of stellar populations in spheroids based on individual stellar abundances and kinematics. The observed properties point to a very complex nature, which is hard to extrapolate from a few fields.}
{We present a method to obtain reddening maps and to trace structure and metallicity gradients of the bulge using data from the recently started ESO public survey Vista Variables in the Via Lactea (VVV). The method is used to derive properties of the fields along the minor axis.}
{We derive the mean $J-K_s$ color of the red clump (RC) giants in 1835 subfields in the Bulge region with $-8^\circ<b<-0.4^\circ$ and $0.2^\circ<l<1.7^\circ$, and compare it to the color of RC stars in Baade's Window for which we adopt $E(B-V)=0.55$. This allows us to derive the reddening map on a small enough scale to minimize the problems arising from differential extinction. The dereddened magnitudes are then used to build the bulge luminosity function in regions of $\sim0.4^\circ \times 0.4^\circ$ to obtain the mean RC magnitudes. These are used as distance indicator in order to trace the bulge structure. Finally, for each subfield the derived distance and extinction values have been used to obtain photometric metallicities through interpolation of red giant branch colors on a set of empirical ridge lines. The photometric metallicity distributions are  compared to metallicity distributions obtained from high resolution spectroscopy in the same regions.}
{The reddening determination is sensitive to small scale variations which are clearly visible in our maps. Our results are in agreement within the errors with literature values based on different methods, although our maps have much higher resolution and more complete coverage. The luminosity function clearly shows the double RC recently discovered in 2MASS and OGLE III datasets, hence allowing to trace the X-shape morphology of the bulge. Finally, the mean of the derived photometric metallicity distributions are in remarkable agreement with those obtained from spectroscopy.}
{The VVV survey presents a unique tool to map the bulge properties by means of the consistent method presented here. The remarkable agreement between our results and those presented in literature for the minor axis allows us to safely extend our method to the whole region covered by the survey.}
             
\authorrunning{Gonzalez et al.}
\titlerunning{Reddening and metallicity maps of the Milky Way Bulge}

\maketitle

%

\section{Introduction}

Several studies have attempted to address the problem of the origin
of the Galactic bulge, trying to reconstruct its star formation history
through the analysis of both stellar ages and chemical abundances \citep[][]{mcwilliam94,zoccali03,fulbright07,rich_origlia07,lecureur07,zoccali08,clarkson08,brown10,johnson11,gonzalez11}.
Early studies were restricted to a low extinction window at a latitude $b=-4^\circ$ along the minor axis of the bulge, the so-called Baade's Window. However several recent papers presented the evidence for a high complexity of the bulge structure and stellar population content: i) The Milky Way bulge has a radial metallicity gradient \citep[][]{zoccali08,johnson11} most likely flattening in 
the inner regions \citep[][]{rich_origlia07}; ii) it seems to host two kinematically 
and chemically distinct components \citep[][]{babusiaux10,gonzalez11}; and iii) it has an X-shape 
\citep[][]{mcwilliam10,nataf10}. All these observations strongly argue against the assumption
that the properties of the Galactic bulge stellar population in a few low extinction
windows could be extrapolated to the whole bulge. Further analysis of the chemical properties of stars in other bulge regions, away from the minor axis, has been partially obtained from the latest sample of microlensed dwarfs from \citet[][]{bensby10}. Although microlensing provides chemical abundances all across the bulge, a larger sample is required in order to verify the presence and extent of gradients. Consequently, at present, a different approach based on a larger coverage with a statistically significant stellar sample is the key to disentangle the general properties of the bulge.

In this context, a base for such larger coverage studies is required in order to analyze and correctly interpret results. In particular, \citet[][]{zoccali03} and later \citet[][]{johnson11} showed how the upper red giant branch photometry can be used to obtain photometric metallicity distributions for a large number of bulge stars in two fields at $b=-6^\circ$ and $b=-8^\circ$, respectively. Using a similar photometric approach to derive the metallicity distribution of a much larger bulge region will provide for the first time a general picture of the iron content of the galactic bulge. Unfortunately, in order to reach this goal a first challenge arises from the extinction variations across different bulge sightlines. While most of the studies relied on the large scale extinction maps of \citet[][]{schlegel}, it is well known that these maps become unreliable when inner regions of the bulge closer to the galactic plane ($|b|<3$) are considered. \citet[][]{schultheis99}, based on DENIS photometry, and \citet[][]{dutra03}, based on 2MASS, used the red giant branch photometric properties to obtain 2D extinction maps in several inner bulge regions. \citet[][]{marshall06} presented 3D models of the extinction properties of the inner Galaxy with a resolution of 15'. However, at this point a homogeneous extinction map of the bulge including both the outer and the most inner regions, where high resolution is required to avoid differential extinction, is still necessary.
 
Regarding the structure of the bulge, the work of \citet[][]{nataf10} and \citet[][]{mcwilliam10}, based on OGLE and 2MASS photometry respectively, provided hints for a X-shape morphology in the outer regions of the bulge which however, seems to merge into a single structure in the innermost regions, i.e. the bar \citep[][]{rattenbury07}. Both \citet{mcwilliam10} and \citet{nataf10} used the observed properties of the RC stars at different lines of sight, which unfortunately restricted the studies to the outer bulge regions due to the non-complete optical coverage of OGLE and the relatively bright limit magnitude of 2MASS. For a thorough study of the bulge structure, and to homogeneously trace both the inner and outer regions, high spatial coverage and deep near-IR photometry are necessary.

Both reddening and structural properties can be obtained from the photometric analysis of RC giants across the Galactic bulge. RC giants are core-helium burning stars, equivalent of the horizontal branch stars in old metal-poor populations, and are easily identified in bulge color-magnitude diagrams. Their intrinsic magnitude has a fairly well understood dependence on population properties such as age or metallicity \citep{girardi+salaris01}, and reddening effects in color can be clearly identified. For old and relatively metal-rich population such as found in the Galactic bulge \citep[e.g.][and references therein]{brown10} the population corrections for red clump absolute K-band magnitude are of the order of 0.1~mag \citep{salaris+girardi02}. For this reason, photometric properties of red clump giants can be used to trace simultaneously extinction and relative distances in the bulge. The VISTA Variables in the Via Lactea (VVV) Survey, mapping 520 deg$^2$ including the southern disk and bulge of the Milky Way \citep[][]{vvv10}, provides for the first time deep enough near-IR photometry to trace RC stars in all bulge regions including highly reddened innermost parts.

In this paper, we present the first analysis of the Galactic bulge RC giants based on VVV data for a region along the minor axis. We analyze the RC properties in the color magnitude diagrams along different lines of sight to infer reddening properties and mean distances. This information is then coupled with 2MASS photometry of the upper red giant branch (RGB) stars in order to further derive bulge metallicity distributions. This paper describes in detail the method and compares the results with the present literature. In a forthcoming paper we aim to apply the method to the complete 300 deg$^2$ coverage of the VVV bulge area, deriving homogeneous extinction, morphological structure and metallicity maps of the entire Galactic bulge.        

\section{VVV survey data}

\begin{table}
\begin{center}
\caption{Description of the analyzed fields. \label{fields}}
\begin{tabular}{c c c c c c c}
\\[3pt]
\hline
VVV & $l$ & $b$ & Obs. & \multicolumn{3}{c}{seeing (arcsec)} \\
field  & deg & deg & Date &  $J$   &  $H$   & $K_s$ \\
\hline
b236 & 0.93 & -7.60 & 13/08/2010  & 0.90 & 0.80 & 0.79 \\
b250 & 0.91 & -6.51 & 13/08/2010  & 0.93 & 0.83 & 0.88 \\
b264 & 0.89 & -5.41 & 14/08/2010  & 0.91 & 0.83 & 0.81 \\
b278 & 0.88 & -4.32 & 22/04/2010  & 0.91 & 0.86 & 0.84 \\
b292 & 0.87 & -3.23 & 30/08/2010  & 1.20 & 1.09 & 1.13 \\
b306 & 0.86 & -2.14 & 30/08/2010  & 1.41 & 1.10 & 1.14 \\
b320 & 0.85 & -1.04 & 22/04/2010  & 0.81 & 0.76 & 0.71 \\
\hline
\end{tabular}
\end{center}
\end{table}

The VISTA Variables in Via Lactea (VVV)\footnote{http://vvvsurvey.org} is the Galactic near-IR ESO VISTA public survey that started collecting data in 2010. The VVV area covers $\sim520$  deg$^2$ and will be imaged repeatedly in $K_s$-band with the aim to search for variable stars such as RR~Lyrae in the Milky Way bulge, as well as Cepheids and eclipsing binaries across the plane. In addition to the main variability campaign, the survey area is fully imaged in 5 photometric bands: Z, Y, J, H, and $K_s$. 

The VVV survey has two main components: the 300 sq. deg  VVV Galactic bulge survey area covering  
$-10^\circ < l <+10^\circ$ and $-10^\circ < b < 5^\circ$, and the 220 sq. deg VVV plane survey area spanning between  $295^\circ < l <350^\circ$ and $-2^\circ < b < 2^\circ$. In the first observing season in 2010 95\% of the total area has been imaged in J, H and K$_s$ wavebands, while the YZ imaging covered 84\% of the disk and 40\% of the bulge areas. The variability campaign has also started, but mostly it will be carried out in the upcoming observing seasons. The survey is expected to span 5 years of observations, enabling to characterize well both short and long period variable stars in the Milky Way bulge and the plane.

Here we describe only data used in the analysis presented in this paper:  the single-epoch JHKs coverage of the bulge. The more complete description of the whole survey and the observation strategy is available in \citet{vvv10} and some first results from the VVV survey are presented in \citet{saito+10}.

The observations are carried out in service mode with the VIRCAM camera on VISTA 4.1m telescope \citep{emerson+04, emerson+sutherland10} located at ESO Cerro Paranal 
Observatory in Chile on its own peak, some 1.5km away from the Very Large Telescope. 
The bulge survey area consists of 196 tiles, each $1.48 \times 1.18$ sq. deg in size.  VIRCAM focal plane is sparsely populated with $4 \times 4$ Raytheon VIRGO $2048 \times 2048$ pixel arrays, and 6 exposures (so called pawprints) are needed to fully cover the 1.5 square degree field (so called tile). The mean pixel scale is 0.34 arcsec, and the point spread function delivered by the telescope and camera has $FWHM \sim 0.51$~arcsec. The median image quality measured on the reduced VVV tile images ranges between 0.8" for $K_s$, 0.9" for J band, and up to 1.0" for Z-band. Few tiles were observed out of seeing constrains and were later repeated. This is the case for tiles b292 and b306 for which we currently only have reduced images taken with a seeing larger than 1.0". In the future analysis of the whole bulge sample, these tiles will be replaced with newer observations. 

VVV tiles are made combining 12 (two per pawprint) images for $H$, $K_s$ and $J$ band tiles. Each $H$ and $K_s$ image is a single 4 sec exposure, while $J$ band images are sums of two 6 sec exposures. Each pixel in a tile has therefore a minimum of 16 sec exposure in $H$ and $K_s$ and 24 sec in $J$, and is an average of at least 4 individual exposures. VISTA data are processed through the VISTA Data Flow System (VDFS) by the Cambridge Astronomical Survey Unit (CASU). The basic data reduction steps consist of reset correction, dark subtraction, linearity and flat field correction, sky background subtraction, destriping step, jitter stacking and tiling. Finally the astrometric and photometric calibration are applied using 2MASS sources in the images, and the single band catalogues are produced \citep{lewis+10}.  In this work we use the single band tile images and  catalogues version 1.1\footnote{More details about VISTA data processing is available from {\tt \tiny http://casu.ast.cam.ac.uk/surveys-projects/vista/data-processing}}.

The tiles analysed in this paper are located along the minor axis at $l \sim 0^\circ$ and spanning 
$-8^\circ \la b < -0.5^\circ$. Table~\ref{fields} lists center tile galactic coordinates, date of observation, and the FWHM of the PSF in arcsec measured on tiles. Given that data for each tile in three filters $J$, $H$, and $K_s$ are taken consecutively, the airmass is approximately the same for all filters in a tile. 
\begin{figure}
\begin{center}
\includegraphics[scale=0.5,angle=0]{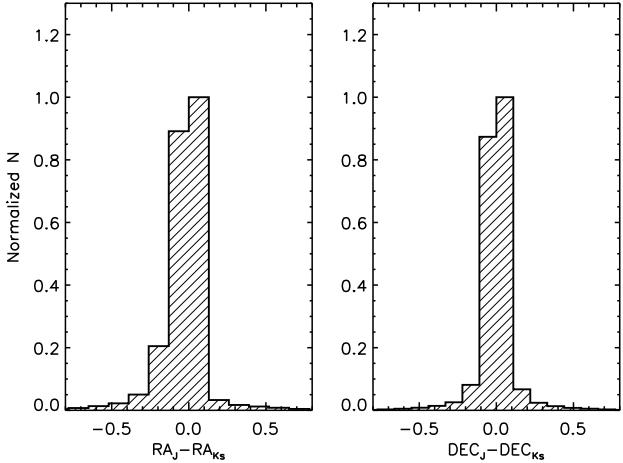}
\caption{Distributions of the RA and DEC differences in arcsec between matched sources of single band, J and Ks, photometric catalogs. Crossmatching for field b278 is shown here as an example}
\label{stilts}
\end{center}
\end{figure}
\subsection{Multiband catalogs}
Target coordinates and $J$,$H$,$K_s$ magnitudes were extracted from the single band catalogs produced by the CASU. These catalogs were then matched using the code STILTS \citep[][]{taylor06}. In particular, the routine \textit{tskymatch2} is used to perform a match based on the proximity of sky positions of stars in each catalog. Several tests were carried out by modifying the maximum separation radius to match detections between $J$, $H$ and $K_s$ catalogs. The majority of stars were always found within a separation smaller than 0.5", in good agreement with the astrometric precision expected from CASU pipeline. Therefore, a maximum allowed separation of 1" was used to ensure that our matched catalogs contain most of the sources without risk of including wrong matches at a larger separation, and at the same time we still leave the possibility for some minor astrometry distortions in the large VIRCAM field of view. Figure~\ref{stilts} shows the distributions of the RA and DEC difference for J and Ks catalogue matches in the VVV field b278. 

The single band tile catalogs produced by CASU contain a flag which identifies detections as stellar or non-stellar sources. We decided to only use sources with the stellar flag in all three bands. Although this cleaning procedure implies that only $\sim$75\% of the original number of detections remain in the final multiband catalog, it allows us to work with as clean as possible sample of bulge stars, which is ideal for our purpose, and it does not bias our results.

\subsection{Photometric calibration}

\begin{figure}
\begin{center}
\includegraphics[scale=0.54,angle=0]{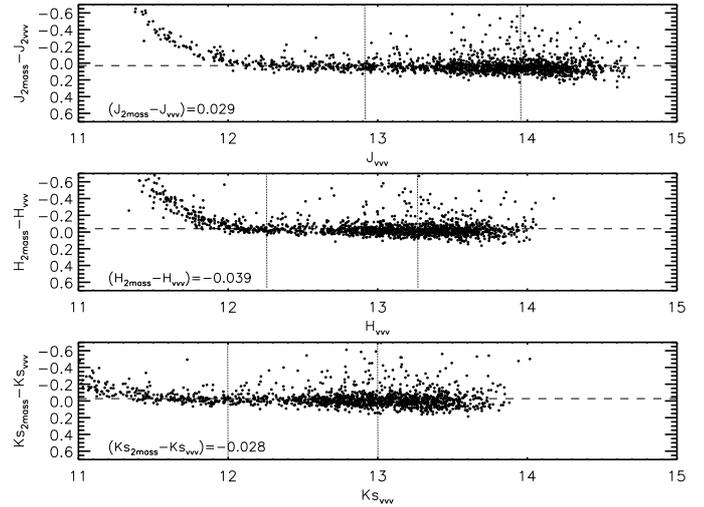}
\caption{$J$,$H$ and $K_s$ magnitudes differences between VVV catalogs and 2MASS. Dashed lines show the mean difference between both catalogs in the range denoted by the solid lines}
\label{2mass}
\end{center}
\end{figure}

The VVV images extend about 4 mag deeper than 2MASS\footnote{The exact difference between VVV and 2MASS photometry depth depends strongly on the stellar density in the field.} in spite of the short exposure times because of the difference in telescope apertures and spatial resolution. However, the larger telescope aperture of VISTA results in saturation for brightest stars. For this reason 2MASS photometry is required to complete the bright end of the VVV catalogs. Figure~\ref{2mass} shows the $J$, $H$ and $K_s$ magnitude differences between VVV and 2MASS sources in a common 10x10 arcmin window within the tile b278. We only adopted 2MASS sources with best photometric quality (flag AAA). Two important points can be observed from Fig.~\ref{2mass}: i) The saturation magnitude in the VVV photometry is clearly observed at 
$K_{s,\rm{vvv}}<12$; and ii) there is a very small offset between the zero points of both catalogs. Still, we must ensure that both catalogues are in the same photometric system to be able to complete the saturated bright end of the VVV photometry with 2MASS. In doing this, only stars common to both surveys and with $13>K_{s,\rm{vvv}}>12$ (see Fig.2) were used to bring the VVV catalogues onto the 2MASS photometric system. The magnitude range selected for the calibration ensures that the differences are not affected by larger errors in the photometry for 2MASS or saturation for VVV measurements. This procedure was carried on independently for each VVV tile and zero points were calculated for each one of the catalogs. Once calibrated, stars brighter than $K_s\sim12$ from 2MASS were used to complete the bright end of our final VVV catalog. These catalogs, corrected for saturation and calibrated to 2MASS zero points were used in further analysis. Hence, hereafter we refer to the usual $J$,$H$,$K_s$ magnitudes corresponding to the 2MASS photometric system. 

\subsection{Bulge minor axis color-magnitude diagram}

\begin{figure}
\begin{center}
\includegraphics[scale=0.65,angle=0]{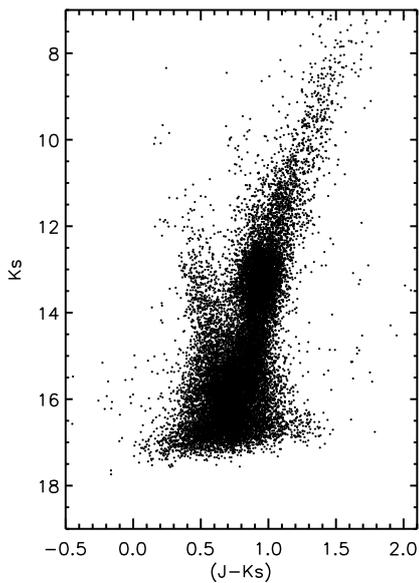}
\caption{VVV observed ($J-K_s$,$K_s$) CMD of a  $20' \times 20'$ region within the tile b278 calibrated and completed with 2MASS photometry.}
\label{cmd_final}
\end{center}
\end{figure}

As an example, Figure~\ref{cmd_final} shows the $K_s$, $J-K_s$ color-magnitude diagram (CMD) of a selected region in the field b278. The photometry samples the entire RGB from the base up to the tip ($K_s\sim8$). The RGB is quite bent as expected for metal-rich populations and its relatively large width is due to metallicity spread \citep{zoccali03}, depth effects and some differential reddening. The red clump is located at $K_s\sim13$, and it is partially merged with the RGB bump. Although the photometry is deep enough to properly sample the sub giant branch, the almost vertical bluer sequence due to foreground disk stars prevents to detect its exact location.

\section{Relative reddening values}

The extinction corrections are derived by using a differential method based on the comparison between the mean color of the RC stars observed in all fields with those of a reference field with well known extinction, the Baade's Window field. The important assumption we are taking here is that the differences in terms of average age and metallicity of the bulge stellar populations can be accounted for with only a small correction on the color of the RC stars in the bulge fields, given the old age and nearly solar metallicity of the Bulge and the presence of a known metallicity gradient \citep{zoccali08}. 

\begin{figure}
\begin{center}
\includegraphics[scale=0.50,angle=0]{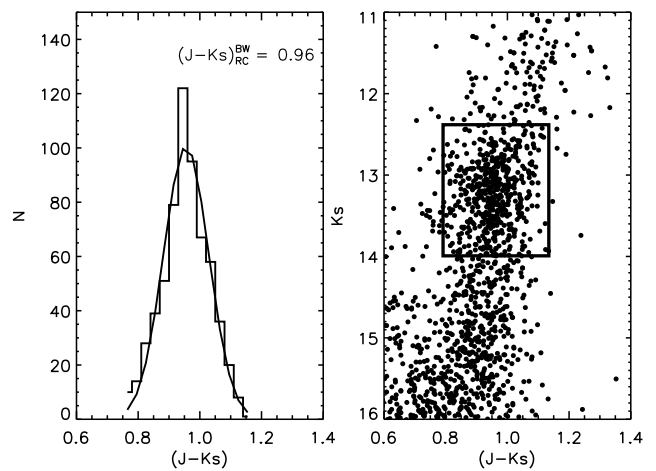}
\caption{Right panel shows the observed color magnitude diagram for a $10 \times 10$ arcmin region centered in Baade's Window and a selection box corresponding to the red clump. Left panel shows the color distribution of these red clump stars. Overplotted is the best fit Gaussian curve with the mean RC color.}
\label{color}
\end{center}
\end{figure}

\begin{figure}
\begin{center}
\includegraphics[scale=0.50,angle=0]{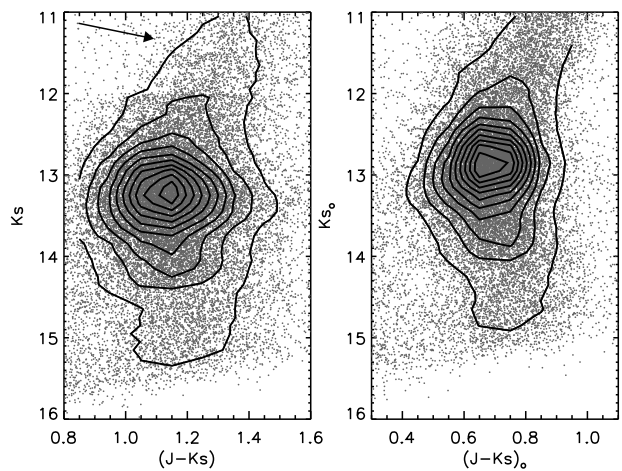}
\caption{Red clump region of the observed color magnitude diagram for tile b306 (right panel) compared to the de-reddened one obtained by our procedure (left panel). Only a subsample of the stars are plotted in each CMD and contour levels are shown to mark the clump position and shape. The arrow in the right panel shows the reddening vector for a change in E(B-V) of 0.5 magnitudes.}
\label{cmd_red}
\end{center}
\end{figure}

In order to obtain the properties of a reference RC population we have analyzed one of the fields for which we have spectroscopically determined iron and alpha element abundances \citep[][]{zoccali08, gonzalez11}. We have measured the color of the red clump in a $10' \times 10'$  region centered in $l=1.14$ and $b=-4.18$ (hereafter Baade's Window) for which we have adopted a reddening value of E(B-V)=0.55 as in \citet[][]{zoccali08}. Figure~\ref{color} shows the observed CMD for this region and the ($J-K_s$) color histogram for which we find a mean value of ($J-K_s$)$_{RC}^{BW}$=0.96 based on a Gaussian fit to the color distribution. Adopting this same procedure, the extinction E(B-V) of a bulge field at (l,b) can be related to that of Baade's Window with the following equation:

\begin{equation}
 E(B-V)=E(B-V)_{BW}-\Delta(J-K_s)_{RC}/(0.87-0.35)
\end{equation} 
where $E(B-V)_{BW}$ is the extinction in Baade's window and $\Delta(J-K_s)_{RC}$ is the difference between the color of the RC in Baade's Window, ($J-K_s$)$_{RC}^{BW}$, and the observed color of the RC for the field at (l,b), ($J-K_s$)$_{RC}^{lb}$, as obtained from the Gaussian fit. For our procedure we adopted the reddening law from \citet[][]{cardelli89}: $A_K=0.35\cdot E(B-V)$, $A_H=0.59\cdot E(B-V)$ and $A_J=0.87\cdot E(B-V)$. We note that different reddening laws have been derived for some inner bulge fields ($|b|<2$) \citep[][and references therein]{nishiyama09}, however these do not cover the full area that we are investigating. On the other hand, the bulge RR Lyrae study from \citet[][]{kunder08} showed that, although variations are observed in particular directions, a standard reddening law of $R_V\sim3.1$ is, on average, valid for bulge studies at larger distances from the Galactic plane. Still, the adoption of a different reddening law, such as the one in \citet[][]{nishiyama09} where $A_K=0.528\cdot E(J-K)$ instead of the standard $A_K=0.640\cdot E(J-K)$ would affect our results by less than 0.06 mags in $K_s$ band, for the range of extinctions observed in the regions where we provide a bulge stellar population analysis.

We are able to use this technique in all bulge fields covered by the survey as VVV photometry reaches the RC even in the highly extincted regions. We have divided each VVV field into smaller subfields of $10' \times 10'$ for ($b<-5^\circ$), $6' \times 6'$ for ($-5^\circ<b<-1.5^\circ$) and $3' \times 3'$ for ($-1.5^\circ<b<0^\circ$) in order to minimize the effects of differential reddening and at the same time, to accurately trace the RC with robust number statistics. For fields at larger distances from the plane ($b<-5^\circ$) the density of stars and the differential reddening variations are smaller while the opposite occurs in regions closer to the plane. For this reason we use sub-fields with slightly different sizes and derive a single reddening value for each of the sub-fields.

A fully de-reddened CMD zoomed on the RC region is shown in Fig~\ref{cmd_red} for tile b306. The effect of differential reddening is clearly observable as an increase of the RC width (Fig.~\ref{cmd_red}, right panel), and it is consistent with the reddening vector shown in the figure to indicate the effect of  E(B-V) difference of 0.5 mag. We have carried out the same procedure in all 6 tiles along the bulge minor axis (Table~\ref{fields}). The main objective was to obtain a homogeneously de-reddened dataset of $J$,$H$,$K_s$ photometry to investigate RC properties and metallicities, however at the same time we also got a consistent reddening map for this bulge region.

\subsection{Minor axis reddening map}

\begin{figure}
\begin{center}
\includegraphics[scale=0.95]{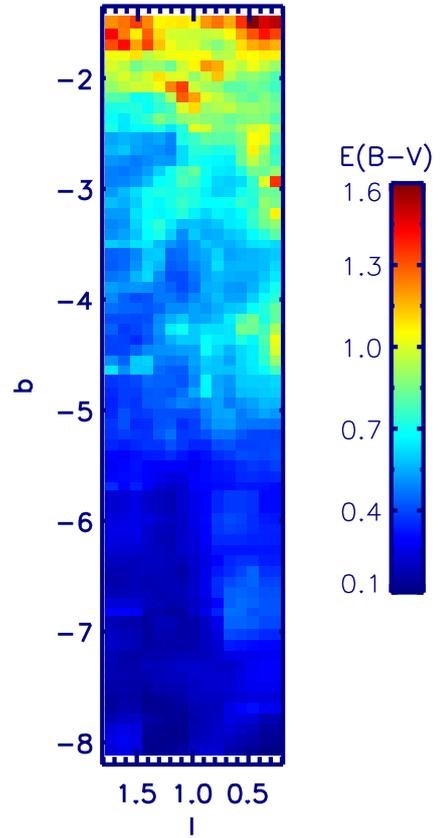}
\caption{Reddening map along the minor axis of the bulge. Regions with $b<-5^\circ$ are smoothed to 6' boxes for visualization.}
\label{red}
\end{center}
\end{figure}

\begin{figure}
\begin{center}
\includegraphics[scale=0.75,angle=0]{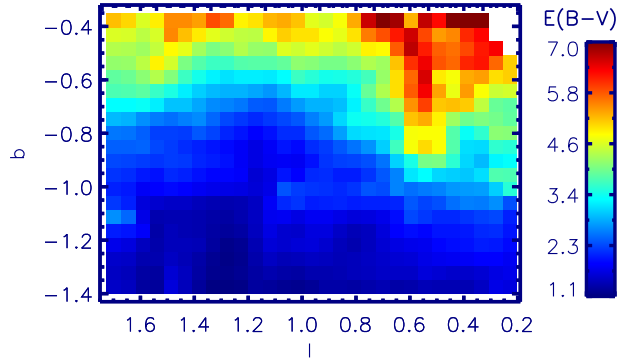}
\caption{Reddening map for the inner regions of the bulge along the minor axis.}
\label{red_in}
\end{center}
\end{figure}

Given the density of stars in the inner fields along the minor axis, the area was split into 1835 subfields with boxes of 3' on a side for fields within inner 1.5$^\circ$, 6' for $-5<b<-1.5^\circ$ and 10' for $b<-5^\circ$. Our map is therefore in-sensitive to reddening variations on smaller scales. Figure~\ref{red} shows the reddening map obtained for the bulge minor axis between $b=-8^\circ$ and $b=-1.5^\circ$ at longitudes  between $0^\circ$ and $2^\circ$. As expected, although the small scale variations are clearly observed, they become particularly strong in the inner regions as can be observed in Figure~\ref{red_in} which shows the reddening map for $b>-1.4$. This indicates that the use of lower resolution maps such as the one of \citet[][]{marshall06} (15') for the very inner bulge regions, near to the galactic plane, will not be sensitive to such variations.
 
\begin{figure}
\begin{center}
\includegraphics[scale=0.5,angle=0]{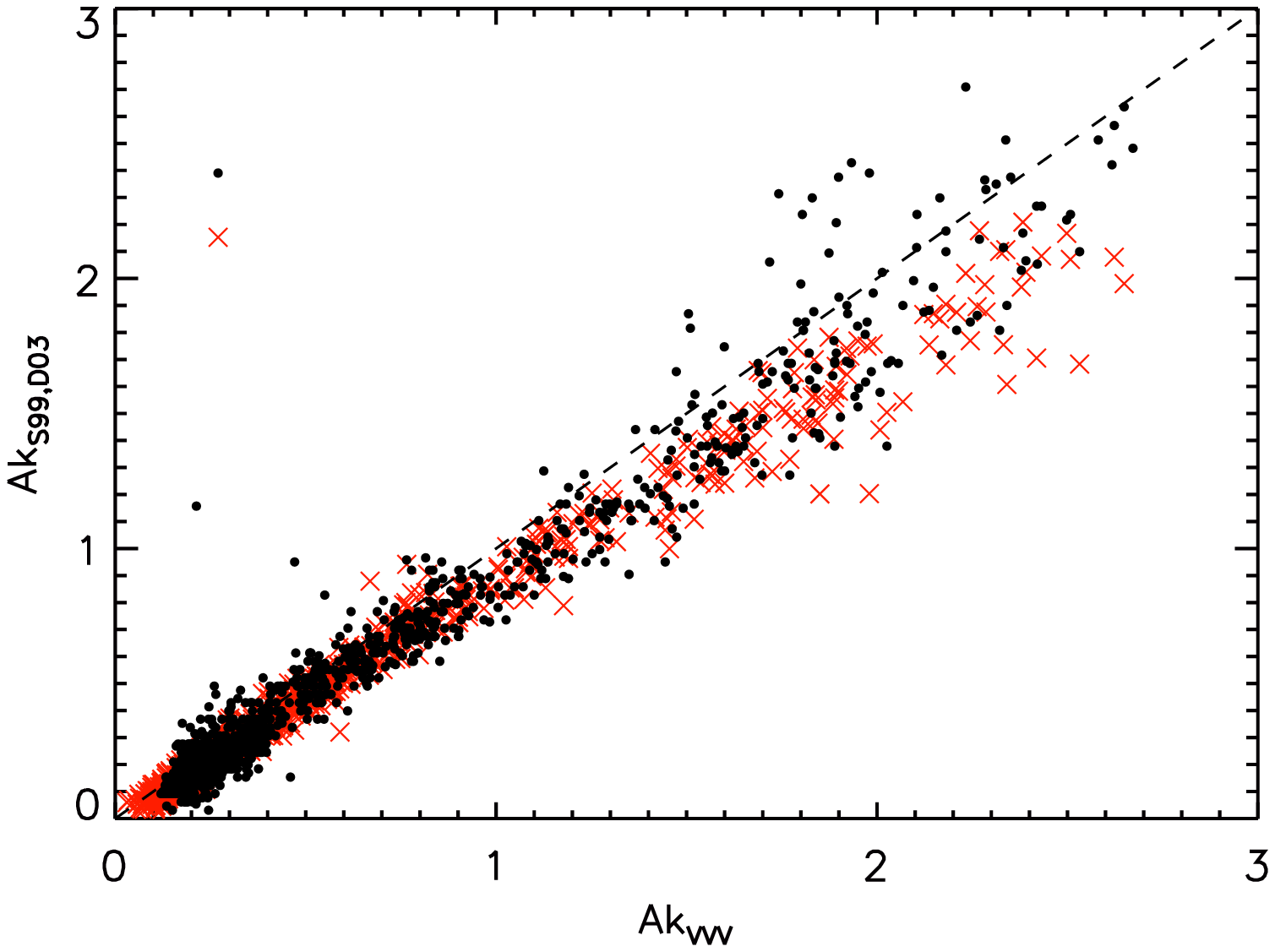}
\caption{Comparison between the $A_k$ values obtained in our work and those of \citet[][black filled circles]{schultheis99} and \citet[][red crosses]{dutra03}}
\label{red_comp}
\end{center}
\end{figure}

High resolution extinction maps for the galactic bulge, including the inner regions, have been obtained by \citet[][]{schultheis99} and \citet[][]{dutra03}. \citet[][]{schultheis99} used \citet[][]{bertelli94} isochrones fitting to red giant branch stars in DENIS ($J$,$H$,$K_s$) CMDs built in subfields of $2 \times 2$ arcmin. Their reference isochrones where calculated by assuming a 10 Gyr solar metallicity bulge population at 8 kpc from the Sun. On the other hand, \citet[][]{dutra03} determined reddening with a similar procedure based on 2MASS photometry, but using a composite observed bulge CMD as reference instead of adopting an isochrone model. 

Figure~\ref{red_comp} shows the difference between our reddening values and those from \citet[][]{schultheis99} and \citet[][]{dutra03}. The comparison was made by matching the central coordinates of the subfields of the different studies. We have applied \citet[][]{cardelli89} extinction transformations to calculate $A_K$ values from these studies, in which other transformations were adopted, in order to obtain a consistent comparison with our results. We systematically find larger reddening values and the difference becomes particularly conspicuous in high extinction regions. This can be caused by the different assumptions that each technique adopted. Our map is relative to the zero point adopted from the observed ($J-K_s$) color of the RC in a $10' \times 10'$ field at Baade's Window, for which we assume a reddening value of E(B-V)=0.55. This alone could introduce a systematic offset depending on this particular reddening value. Also, our method assumes that the intrinsic color of the RC is, in the mean, the same all across the Bulge. Unfortunately, this or similar assumptions are always necessary. In particular, the studies of \citet[][]{schultheis99} and \citet[][]{dutra03} are also based on assumptions of the homogeneity of stellar populations in the bulge and refer to particular mean values in age and metallicity which might introduce the observed offsets. However, RC properties used in our work have a smaller dependence on metallicity, within the known bulge values, in comparisson to those of the RGB. Altogether, when considering the difference in the methods and datasets, the agreement between the three studies is remarkable up to $A_K\sim 1$. For higher extinctions, given that those regions are also the ones suffering higher crowding problems, we expect that the high resolution and deeper photometry of VISTA provide the most reliable results.

\section{Bulge luminosity function and distances}

With the de-reddened photometry of VVV obtained as explained in the previous section, we are able to use the RC magnitude as a distance indicator \citep{paczynski+stanek98, alves+02, grocholski+sarajedini02}. We divided each VVV field in four subfields of 
$\sim 0.4^\circ \times 0.4^\circ$ in which we built the luminosity function (LF) and used it to obtain the mean observed magnitude of the RC, $K_{s_0}^{RC}$,  by means of a Gaussian fitting procedure \citep[e.g.][]{stanek98,babusiaux05} following the function:
\begin{equation}
N(K_{s_0})=a+bK_{s_0}+cK_{s_0}^2+\frac{N_{RC}} {\sigma_{RC}\sqrt{2\pi}}\exp\Big[\frac{(K_{s_0}^{RC}-K_{s_0})^2}{2\sigma^2_{RC}}\Big]
\end{equation}

\begin{figure}
\begin{center}
\includegraphics[scale=0.50,angle=0]{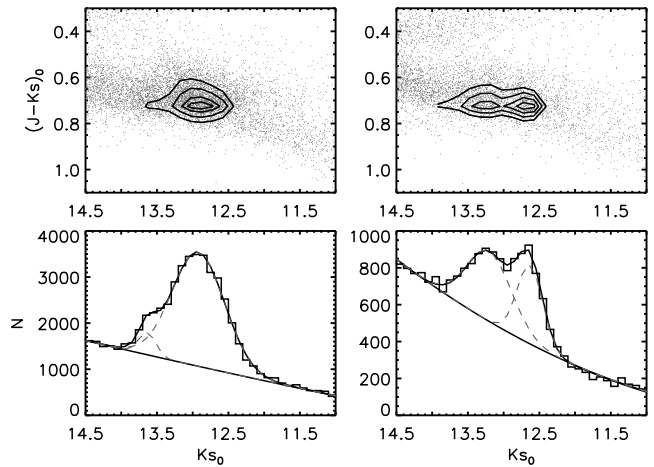}
\caption{Lower panels show the luminosity functions for a subfield in Baade's Window and at b=-6. The underlying RGB is fitted with a second order polynomial and the red clump is fitted with two Gaussians centered at $K_{s_0}=12.9$ and $K_{s_0}=13.2$ for b=-6 and with a single Gaussian centered at $K_{s_0}=12.9$ for Baade's Window. Upper panels show the corresponding CMD for each field, oriented with color along the y-axis and magnitude along the x-axis. Density contours denote the single and double RC in each field.}
\label{LF}
\end{center}
\end{figure}

We first applied a color cut at $(J-K_s)>0.5$ to minimize the contamination from foreground stars and then fitted the underlying RGB LF using a second order polinomial corresponding to the first 3 terms of Eq. (2). The RC was then fitted with two Gaussians, in the form of the forth term in Eq. (2), in the regions where the two red clumps are observed, i.e for latitudes $b<-5^\circ $ \citep[][]{mcwilliam10} and with a single Gaussian in the inner regions. An additional Gaussian fit was performed when the red giant branch bump (RGBb) was detected. Upper panels of Fig.\ref{LF} show the final de-reddened CMDs for Baade's Window and the field at $b=-6^\circ$ where the density contours show clearly the single and double RC features in each field. The lower panels of Fig. \ref{LF} show the de-reddened $K_{s_0}$ band luminosity functions for both fields with overplotted polynomial + Gaussian fits. The polynomial function fits very well the underlying red giant branch population, while Gaussian functions are required to fit RCs and RGBb. A single RC is observed at $K_{s_0}\sim 12.9$ in $b=-4^\circ$, while at $b=-6^\circ$ the RC region shows two components at $K_{s_0}\sim13.2$ and $K_{s_0}\sim12.6$. Additionally, the RGBb is observed at $K_{s_0}\sim13.6$ in $b=-4^\circ$. The latter is in agreement with the measurements of the bulge RGBb by \citet[][]{nataf11} using OGLE photometry.

We use the observed mean RC magnitudes in order to obtain distances for each of our subfields through comparison of the measured values with the intrinsic RC magnitude for a 10 Gyr solar metallicity population of M$_{K_s}=-1.55$ based on \citet[][]{pietri04} Teramo stellar evolution models \citep[see also][]{vanhel07}.

As we already discussed in the case of the reddening calculations, the assumption of homogeneous stellar populations on the bulge are the main source of uncertainties. The intrinsic magnitude of the RC is known to have a dependence on age and metallicity \citep{salaris+girardi02, pietrzynski+03}. While the age of the bulge is expected to be homogeneously old ($\sim$10 Gyr), the metallicity is known to change at least along the minor axis \citep[][]{zoccali08}. According to \citet[][]{zoccali08}, the Bulge metal content varies from [Fe/H]$\sim-0.3$ at $b=-12^\circ$ to [Fe/H]$\sim-0.1$ at $b=-4^\circ$ and there is a hint for a lack of gradients on the inner regions \citep[][]{rich_origlia07}. Teramo stellar evolution models indicate that a 0.2 dex variation in metallicity would affect the intrinsic magnitude of the RC by $\sim$0.10 mag, consistent as well with \citet[][]{salaris+girardi02}. If for a solar metallicity population at 8 kpc from the Sun we assume a main metallicity 0.20 dex lower, our method will result in measurement of distance of $\sim$8.2 kpc instead. Therefore, although this introduces a small error into our calculations we are still able to trace the large scale structure of the bulge.


\citet[][]{mcwilliam10} and \citet[][]{nataf10} showed for the first time the presence of a double RC observed at large latitudes. \citet[][]{mcwilliam10} argued that this feature could be consequence of a X-shaped Bulge. Furthermore, \citet[][]{saito11} mapped the density of RC stars across the bulge using 2MASS photometry and concluded that the X-shape morphology merges to the bar in the inner regions. Our results are in agreement with the measurements in previous studies, showing the clear double peak in the LF of the bulge as can be seen in Fig.\ref{LF}. Our method can be then used to map the innermost Bulge regions which are not considered in the work of \citet[][]{saito11} because of the limiting depth and blending in 2MASS photometry in these inner regions. This analysis is underway and will be the subject of a dedicated paper.

\section{Bulge photometric metallicity distributions}

\begin{figure*}
\begin{center}
\includegraphics[angle=0]{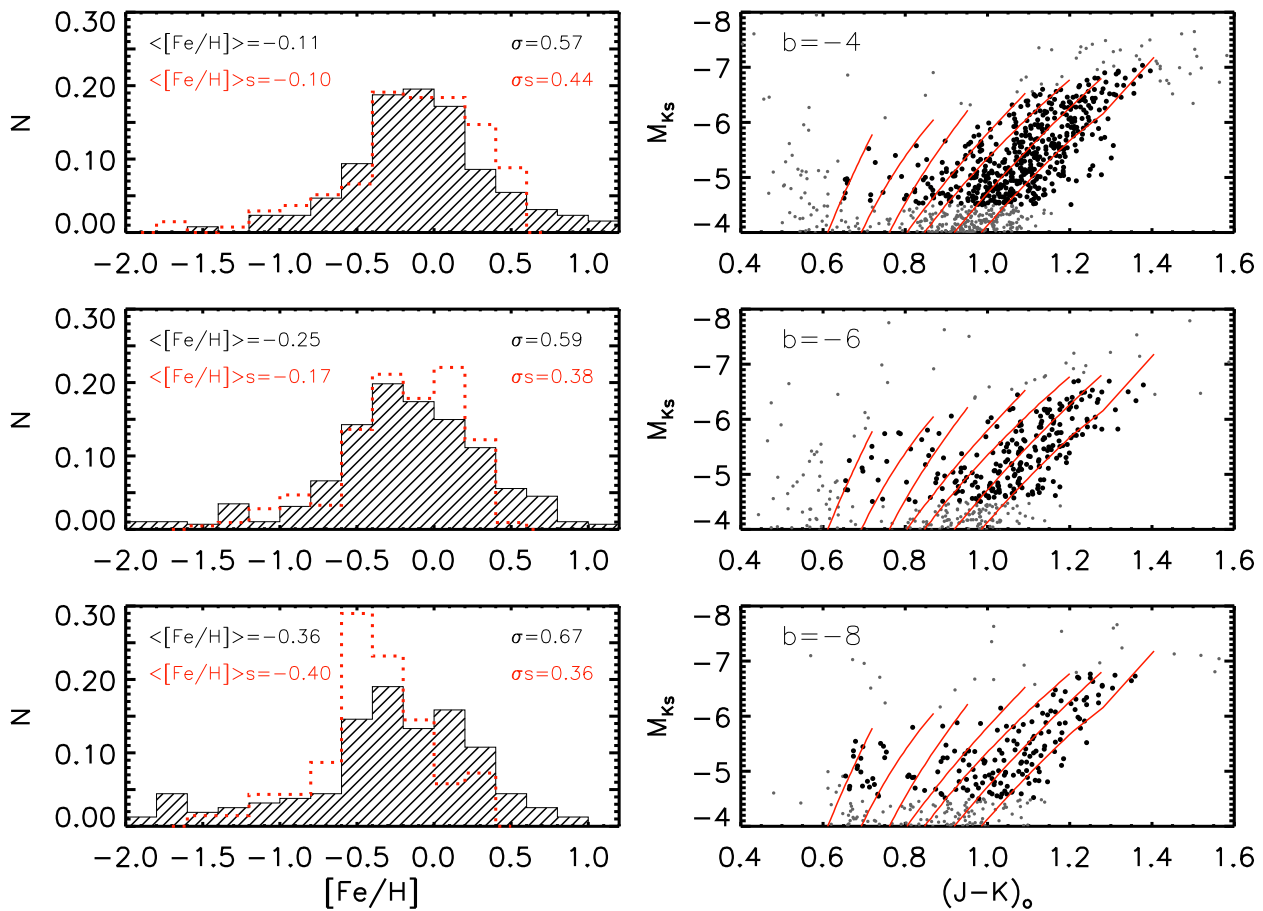}
\caption{{\it Left panels:} Comparison of the photometric (hashed histogram) and spectroscopic (red dotted histogram) MDFs along the Bulge minor axis at $b=-4^\circ$, $-6^\circ$ and $-8^\circ$. The average [Fe/H] and dispersion $\sigma_s$ are also shown for all MDFs. {\it Right panels:}  CMDs in the absolute plane of the fields located along the minor axis at $b=-4^\circ$, $-6^\circ$ and $-8^\circ$ with overplotted the empirical RGB templates. Thick black circles refer to the stars used to derive the photometric MDFs.}
\label{mdf}
\end{center}
\end{figure*}

The final goal of our work is to use the information obtained in the previous sections such as reddening and mean distance to study stellar population metallicity gradients in the bulge. We compare the derived field CMDs with an empirical grid of RGB ridge lines of globular clusters in the [M$_{K_s}$, ($J-K_s$)$_0$] absolute plane. The empirical templates of Galactic halo and bulge globular clusters (M92, M55 , NGC 6752, NGC 362, M 69, NGC 6440, NGC 6528 and NGC 6791), with well know metallicity, is selected from the sample of \citet[][]{valenti04}. We then derive a metallicity value for each field RGB star from its ($J-K_s$)$_0$ color interpolating within the empirical grid. This procedure has been proven to be effective in bulge globular clusters \citep[][]{valenti07}, field \citep[][]{zoccali03,johnson11} as well as in studies of more distant galaxies \citep[e.g.][]{rejkuba+05,lianou+11}.

To transform the observed field CMDs into the absolute plane we use the distance and reddening values derived in the previous sections, while for the cluster ridge lines we adopt the estimates listed in Table 3 of \citet[][]{valenti04}. It is worth to mention that the cluster templates are derived from a homogeneous photometric database calibrated onto the 2MASS photometric system \citep[see][for more details]{valenti04} and they have been selected in order to span a large metallicity range, between -2.16 and +0.35. In particular, we have included in our empirical grid the RGB ridge line of NGC6791 which is one of the oldest \citep[age 6-12Gyr,][and references therein]{carney05} and metal-rich ([Fe/H]=0.35, \citep[][and references therein]{origlia06} open cluster. Furthermore, we notice that the alpha-abundance patterns of the selected templates - enhanced for metallicity up to that of NGC6528 and solar for higher metallicities - match very well the abundance patterns observed in the Bulge fields \citep[][]{gonzalez11, alves10, lecureur07, fulbright07}. This implies that there is no need to adjust the fiducial ridge line of the clusters to match the alpha-abundances of the fields. In the interpolation, stars brighter than M$_{K_s}=-4.5$ were used to include only the RGB region most sensitive to metallicity variations, and to avoid the use of red clump stars \citep[See][for a detailed discussion]{zoccali03}. Additionally, a color cut was applied including only stars with ($J-K_s$)$_0>0.6$ to minimize the disk contamination.

Although this method has been shown to work well in specific cases (i.e. single stellar population and distant galaxies), here a problem may arise from the assumed distance modulus along different latitudes. First, as discussed in the previous sections, the distance estimates are derived assuming an intrinsic RC magnitude which depends on the metallicity. However, the expected error of 0.10 mag for RC stars due to metallicity gradient has small effect on our photometric metallicity measurements because these are based on RGB morphology which has significantly larger sensitivity to metallicity variations. To show that we compare in the next section our photometrically determined metallicity distributions with iron abundances from high resolution spectroscopy.

A second and more important problem arises from the observed double red clump in the luminosity function, which traces the X shape of the bulge. The choice of the bright or the faint RC as the observed value to obtain the absolute magnitudes has a strong impact on the results because it implies a change of up to 1.0 mag along the minor axis. Clearly, whichever RC we assume to be the dominating population, RGB stars from the population corresponding the other RC will have an error in magnitude as large as the separation of both clumps. Since it is impossible to distinguish the two populations on the RGB, we adopt the weighted average magnitude of the two RCs as the observed magnitude to calculate the distance modulus for the RGB stars in that field. This magnitude was calculated as:
\begin{equation}
 K_{s_0}^{RC}=(K_{s_0}^{RC1} N_{RC1}+K_{s_0}^{RC2} N_{RC2})/(N_{RC1}+N_{RC2})
\end{equation}
where $N_{RC1}$ and $N_{RC2}$ are the number of stars in each clump with mean magnitudes $K_{s_0}^{RC1}$ and $K_{s_0}^{RC2}$ respectively.

Therefore, the errors for metallicities for individual stars obtained using our interpolation method are expected to be significantly larger than those based on high resolution spectroscopy. However, the photometric metallicity method provides two unique benefits: (i) the large number statistics, and (ii) most importantly the large coverage, which allows us to investigate gradients. 

\subsection{Comparison to spectroscopic MDF and the minor axis gradient}

The implications of the errors in the method due to assumptions on the reddening law, on the selection of stars on the RC and the assumptions on population corrections for RC absolute magnitude, are hard to take into account within our calculations rigorously and therefore a direct comparison with a more precise method such as high resolution spectroscopy is particularly useful. Several spectroscopic studies have been carried out in fields along the minor axis of the bulge. In particular the metallicity distributions of \citet[][]{zoccali08} presented the evidence for metallicity gradient between latitudes $-4^\circ$, $-6^\circ$ and $-12^\circ$. The study of \citet[][]{johnson11} provided the metallicity distribution of a field at $b=-8^\circ$, which is consistent with the expectation from the metallicities and gradient observed by \citet[][]{zoccali08}. To check the reliability of our photometric metallicity distribution functions (MDFs), we compare the MDFs from our photometric method in $\sim30' \times 30'$ fields with those from spectroscopic studies of \citet[][]{zoccali08} at $b=-4^\circ$ and $b=-6^\circ$ and of \citet[][]{johnson11} at $b=-8^\circ$.   

\begin{figure}
\begin{center}
\includegraphics[scale=0.75]{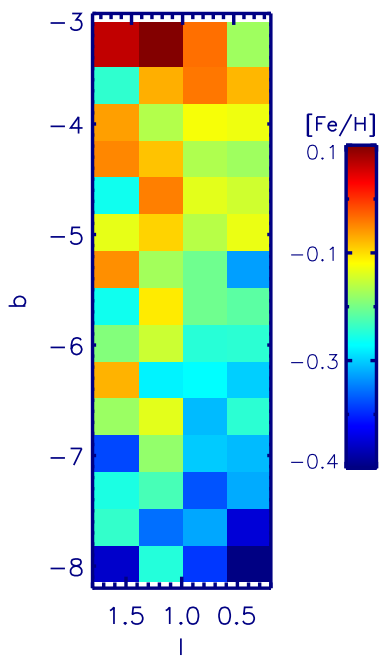}
\caption{A map of the mean values of the metallicity distributions along the bulge minor axis for subfields of $\sim0.4^\circ \times 0.4^\circ$.}
\label{mdf_grad}
\end{center}
\end{figure}

Figure \ref{mdf} shows the remarkable agreement for the mean  photometric (hashed histograms) and spectroscopic (red dotted histograms) MDFs. The metallicity gradient is clearly observed in our data, and the shape of the photometric distributions resemble very well those from spectroscopic studies. The dispersion values in the photometric [Fe/H] distributions are larger than the spectroscopic ones. This may be produced by a combination of factors. In particular, our method relies on extrapolation for stars with metallicities larger than 0.35, the value of the most metal-rich template corresponding to the open cluster NGC6791. This could generate a metal rich tail with no reliable abundances and which might increase the dispersion. On the other hand, the metal-poor tail has a possible disk contamination which could end up also increasing the dispersion of our distributions. However, we do not exclude that such metal-poor stars could actually belong to the bulge. Altogether, we demonstrate that our metallicity determinations are as reliable {\it in the mean} as the high resolution spectroscopic studies. This presents a powerful tool  to trace how metallicity varies not only in discrete fields along the minor axis, but all across the outer bulge. Figure \ref{mdf_grad} shows a map of mean metallicity in bins of $\sim0.4^\circ \times 0.4^\circ$ and it confirms the metallicty gradient along the minor axis. These maps are restricted to $|b|>3^\circ$ as the upper RGB in our dataset corresponds to 2MASS photometry which is limited by crowding at inner regions. In the future paper we will present the metallicity maps of the entire VVV bulge area for $|b|>3^\circ$. 

\section{Conclusions}

We used the VVV survey data to investigate the properties of the Galactic bulge along the minor axis. Within this bulge region there have been several spectroscopic studies which we use to validate our method. The method is based on the use of properties of the red clump giants. The RC color is used to derive the reddening map along the minor axis extending from $-8^\circ <b<-0.5^\circ $. These regions suffer from high extinction and variations on very small scales are observed. Therefore a consistent bulge reddening map obtained from a homogenous tracer all across the bulge, from the innermost crowded and extinct regions to the outer parts is very valuable. Our results for the extinction also show good agreement with maps published by \citet{dutra03}, and \citet{schultheis99}.

We have used the de-reddened magnitudes to build the Ks band luminosity functions and to trace the RC along the minor axis. When using the RC as a distance indicator, we find excellent agreement with previous findings of a double red clump for latitudes larger than $b<-5^\circ$ which trace the X shape of the Galactic bulge \citep[][]{mcwilliam10,nataf10}. We also detect the additional bump in the luminosity function interpreted as the RGBb by \citet[][]{nataf11}. This study will  be extended to the complete 300 deg$^2$ bulge region and will provide a consistent trace of the structure of the bulge including the inner regions. 

Finally, we used the RC distances and reddening map to build the color magnitude diagrams in the absolute plane and then to obtain individual photometric metallicities for RGB stars in the bulge through interpolation on the RGB ridge lines of cluster sample from \citet[][]{valenti04}. The final metallicity distributions were compared to those obtained in the spectroscopic studies of \citet[][]{zoccali08} and \citet[][]{johnson11} showing a remarkable agreement. We demonstrated clearly the possibility to track the observed metallicity gradient along the minor axis. This provides, within the errors of the method, the opportunity to obtain clues for metallicity gradients along other regions of the bulge. Reddening and metallicity maps for the VVV coverage will be released electronically in the survey website.
 

\begin{acknowledgements}
We gratefully acknowledge use of data from the ESO Public Survey
programme ID 179.B-2002 taken with the VISTA telescope, and data
products from the Cambridge Astronomical Survey Unit, and funding from
the FONDAP Center for Astrophysics 15010003, the BASAL CATA Center for
Astrophysics and Associated Technologies PFB-06, the MILENIO Milky Way
Millennium Nucleus from the Ministry of Economycs ICM grant P07-021-F and Proyectos FONDECYT Regular 1087258, 1110393 and 1090213. MZ
is also partially supported by Proyecto Anillo ACT-86. This publication makes use of data products from the Two Micron All Sky Survey, which is a joint project of the University of Massachusetts and Infrared Processing and Analysis Center/California Institute of Technology, funded by the National Aeronautics and Space Administration and the National Science Foundation. We warmly thank the ESO Paranal Observatory staff for performing the observations. We gratefully acknowledge the anonymous referee for helpful comments on this article.
\end{acknowledgements}
\renewcommand*{\bibfont}{\small}
\bibliographystyle{aa}

\bibliography{mybiblio}

\end{document}